\documentclass[aps,floats,twocolumn, preprintnumbers,tightenlines]{revtex4}
\usepackage{epsfig}
\usepackage{amsmath}
\usepackage{amssymb}
\usepackage{amsthm}
\usepackage{graphics}
\usepackage{epsfig}

\newcommand{\bea}{\begin{eqnarray}}
\newcommand{\eea}{\end{eqnarray}}
\newcommand{\beq}{\begin{equation}}
\newcommand{\eeq}{\end{equation}}
\begin{document}

\title{Critical escape velocity of black holes from branes}

\author{Antonino Flachi${}^1$%
\footnote{flachi@yukawa.kyoto-u.ac.jp}, %
Oriol Pujol\`as${}^{2}$%
\footnote{pujolas@ccpp.nyu.edu}, %
Misao Sasaki${}^1$%
\footnote{misao@yukawa.kyoto-u.ac.jp} and %
Takahiro Tanaka${}^3$%
\footnote{tama@scphys.kyoto-u.ac.jp}}%
\address{$^{1}$Yukawa Institute for Theoretical Physics, Kyoto
University, Kyoto 606-8503, Japan\\
$^{2}$Center for Cosmology and
Particle Physics, Department of Physics, New York University, 4
Washington Place, New York, NY
10003 US\\
$^{3}$Department of Physics, Kyoto University, Kyoto
606-8502, Japan} 
\preprint{YITP-06-09}
\preprint{KUNS-2013} 

\begin{abstract}
In recent work we have shown that a black hole stacked on a brane
 escapes once it acquires a recoil velocity. This result was obtained in
 the {\it probe-brane} approximation, {\it i.e.}, when the tension of the
 brane is negligibly small. Therefore, it is not clear whether the
 effect of the brane tension may prevent the black hole from 
 escaping for small recoil velocities. The question is whether a critical escape velocity exists. 
 Here, we analyze 
 this problem by studying the interaction between a Dirac-Nambu-Goto
 brane and a black hole assuming adiabatic (quasi-static)
 evolution.
By describing the brane in a fixed black hole spacetime, which 
restricts our conclusions to lowest order effects in the tension, 
we find that the critical escape velocity does not exist for
 co-dimension one branes, while it does for higher 
co-dimension branes. 
\end{abstract}
\maketitle

\section{Introduction}

A striking prediction of brane world models with TeV Planck scale is
that small black holes can be created in high energy collisions of
particles at energies within the reach of forthcoming experiments
\cite{r1,r1-1,r1-2,r2,r3,r4,r4-b}. When two particles on a brane collide at a center of
mass energy larger than the fundamental Planck scale with 
sufficiently small impact parameter, 
the system collapses and a black hole forms. 
Soon after formation, the black hole will start emitting Hawking
radiation partly into lower dimensional fields localized on the brane,
and partly into higher dimensional bulk modes \cite{r5}. 
The emission of higher dimensional modes will
cause the black hole to recoil into the extra dimensions 
if there is no symmetry that suppresses the recoil velocity, like $Z_2$-symmetry in co-dimension one case. 

The previous simple considerations motivated us to study the interaction between a small
black hole and a brane, paying particular attention to how the
system evolves dynamically when some perturbation gives the black hole a
velocity relative to the brane.
This problem was initially studied in Ref.~\cite{art1}, in the context
of a scalar field model, and it was shown that the black hole is capable
of escaping from the brane when it recoils due to emission of higher
dimensional quanta, but no physical mechanism for the escape was
suggested. This result was obtained in the approximation that the
tension of the brane is negligibly small. 

In Ref.~\cite{art2} we have considered the same problem from the
different perspective of studying the dynamics of Dirac-Nambu-Goto
branes in black hole spacetimes (Relevant results concerning the static case in $4$ dimensions were studied in Ref.~\cite{art2-b}). Our results confirmed the conclusion of
Ref.~\cite{art1} and, in addition, suggested a mechanism for the escape
of the black hole based on the reconnection of the brane. Subsequently,
in Ref.~\cite{art3}, we tested these results within a field theory
model, where the brane was described by a domain wall in a scalar
effective field theory. In this way we could take into account the
recombination processes and fully illustrate the evolution and escape
mechanism. All the previous results were obtained in the approximation
that the tension of the brane has no effect. Although ignoring the
tension is a reasonable assumption when the recoil velocity is large, it
might not be so in the opposite case of small recoil velocity.

In order to make the problem more definite, let us briefly repeat some
relevant points discussed in Ref.~\cite{art2}. There, we investigated
the dynamics of a brane in a fixed Schwarzschild spacetime background
and found that the brane pinches and the black hole slides off the
brane. Having completely ignored the self-gravity of the brane, the
black hole continues to move without being pulled back. Once we take
into account the effects of a finite brane tension, however, it is not
clear if the pinching of the brane occurs before the black hole is
pulled back. When the tension of the brane is large, 
this problem is more difficult to study 
since we need to take account of the deformation of the
geometry caused by the gravity of the brane. Hence, we restrict our
consideration to the effects which are lowest order in the brane
tension next to the probe-brane approximation. 

\begin{figure*}[th]
\scalebox{0.8} {\includegraphics{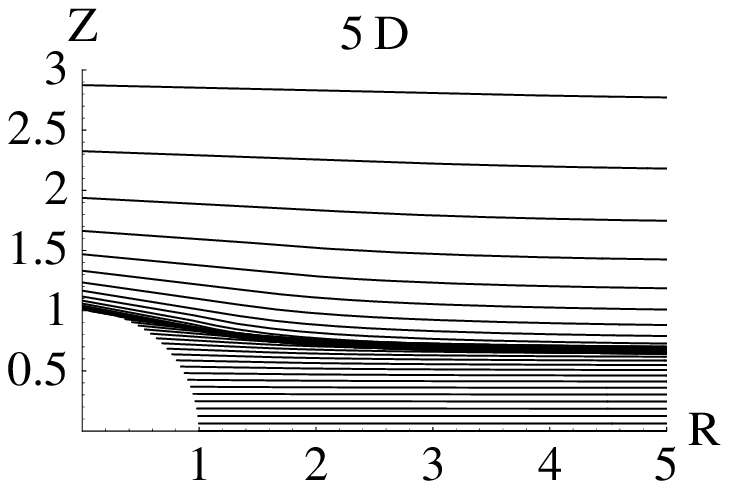}}
\scalebox{0.8} {\includegraphics{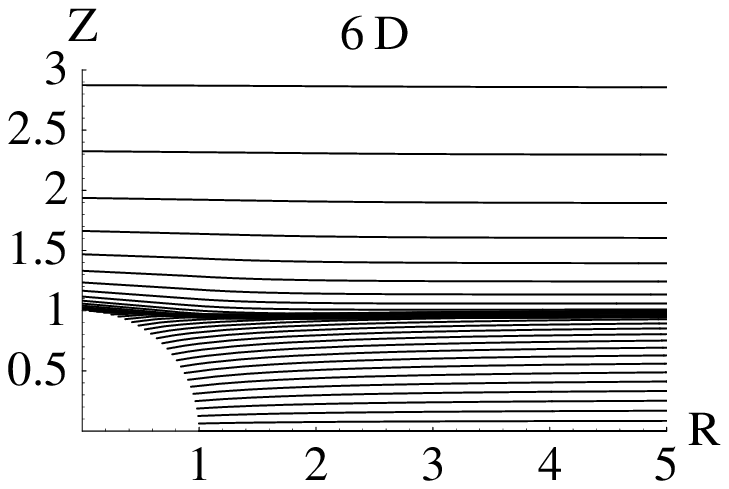}}
\scalebox{0.8} {\includegraphics{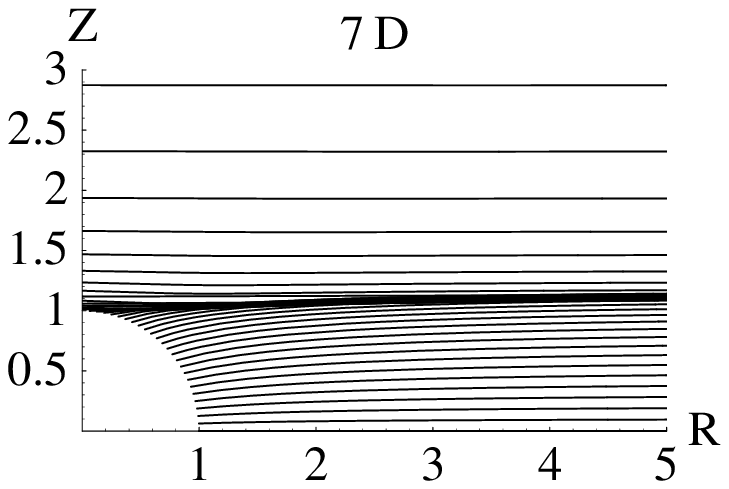}}
\scalebox{0.8} {\includegraphics{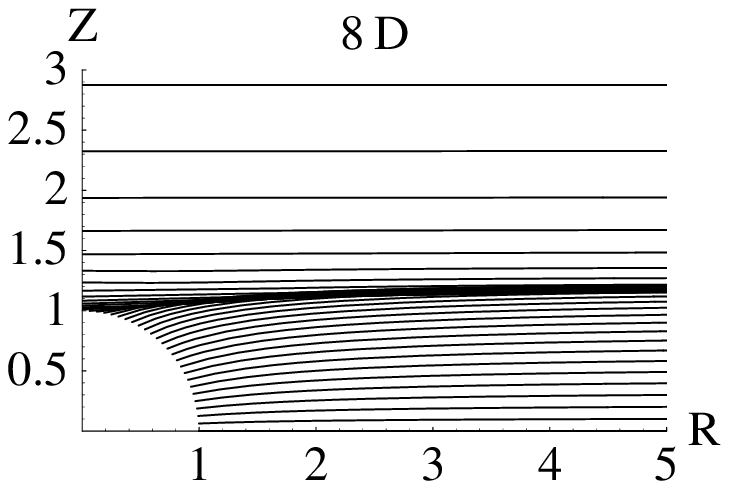}}
\caption{The picture shows the sequence of static configurations
 describing the escape path of the black hole with minimum energy. 
 The plot is obtained by
 solving the equation of motion for the brane numerically in the static
 case. The results reported here refers to the cases with 
 a $d=5, 6, 7, 8$ dimensional bulk
 spacetime and a $p=3$ brane. $R$ and $Z$ are the cylindrical
 coordinates defined in (\ref{cil}).}
\label{fig1}
\end{figure*}

Our strategy goes as follows. We start by considering the quasi-static
evolution of the brane, by which we mean that the brane evolves adiabatically through the sequence of static solutions shown in Fig.~\ref{fig1}. In the next section, we will provide a justification for this. Once the sequence of solutions is fixed, it is
possible to evaluate the energy for each configuration. This
allows one to see how the energy changes along the path that
describes the black hole escape, and to estimate the required initial kinetic energy for the black hole to overcome the height of the energy barrier along the path. 
As one might notice, there are
several non-trivial technical issues to address. 
The most important one concerns the evaluation of the energy 
of an infinitely extended object, which is divergent and thus
requires regularization. This issue will be addressed in the next
section, proposing a regularization scheme. 
The regularization scheme we adopt is to fix the circumferential 
radius of the brane at the outer boundary, 
which is let to infinity in the final steps of the calculation. 
Although it seems unnecessary to justify this regularization 
scheme since it is apparently the most natural way, it is not 
so easy to give a rigorous proof. 
In Appendix we give an argument 
to justify the proposed regularization scheme. 
In Sec.~\ref{sec:Minkowski} we give an analytic comparison of the 
energies between the initial and the final configurations. 
In the initial configuration the brane is 
placed on the equatorial plane of the black hole, while 
in the final one the brane is disconnected and 
far away from the black hole. 
This analysis anticipates the shape of the potential 
along the escape path.  
In Sec.~\ref{sec:barrier} the energy for the sequence of static configurations is evaluated numerically and, 
based on the shape of the potential, 
we discuss the existence of a critical escape velocity.
We find that there is no barrier in the potential for
co-dimension one branes. 
In other words, the force between a black hole and a co-dimension one brane is repulsive. 
Hence, the configuration with a higher dimensional 
black hole attached to such a co-dimension one brane is unstable. 
In contrast, when the co-dimension is equal to or greater 
than two, there
is a potential barrier along the escape path of the black hole 
in the configuration space, and therefore there is a critical escape
velocity. Namely, the configuration with a black hole on a brane is
stable under a perturbation due to a small recoil velocity. 

\section{energy of static configurations}
We consider a system which consists of a black hole and a brane. 
Assuming that the tension of the brane $\sigma$ is small, 
we neglect the gravitational perturbations caused by the presence of 
the brane. Namely, the spacetime is assumed to be given by 
$d$-dimensional Schwarzschild geometry \cite{schw1,schw2}, 
\begin{equation}
 ds^2=-f(r)dt^2+f(r)^{-1}dr^2+r^2d\Omega_{d-2}^2, 
\end{equation}
with $f(r)=1-1/r^{d-3}$, where we set the product of the $d$-dimensional
Newton's constant $G_d$ and the mass of the black hole $m$ to unity. 
We treat the brane as a $(p+1)-$dimensional Dirac-Nambu-Goto 
 membrane. 
Imposing spherical symmetry on the brane, the action is given by 
\begin{eqnarray*}
 S&=&\int dt L\cr
  & \equiv &
  -\tilde{\sigma} \int dt \int dr \left(r\sin\theta\right)^{p-1} 
 \sqrt{1-{r^2 \over f}\dot{\theta}^2
       +r^2 f \theta'^2}~,\nonumber
\end{eqnarray*}
where $\theta$ is the azimuthal inclination angle, $\tilde{\sigma}=\sigma
\Sigma_{p-1}$ and $\Sigma_{p-1}$ is the area of a $(p-1)-$dimensional
unit sphere. 
In the static case, the Lagrangian $L$ is 
identified with the energy $E$ as 
\begin{equation} 
 E =-L
   =\tilde{\sigma}\int dr \left(r\sin\theta\right)^{p-1}
  \left(1+r^2 f \theta'{}^2\right)^{1/2}. 
\end{equation}
It is convenient to describe both $r$ and $\theta$ as functions of a
parameter $\lambda$. Then the energy can be written as 
\beq
  E=\int_0^1 d\lambda\, {\mathfrak L}~,
\eeq
where
\beq
{\mathfrak L} \equiv \tilde{\sigma} \left(r\sin\theta\right)^{p-1} 
  \left(\left({dr\over d\lambda}\right)^2
     +r^2 f \left({d\theta\over d\lambda}\right)^2\right)^{1/2}~.  
\eeq
Using the re-parametrisation invariance of the above expression, we
fixed the boundary values of the parameter $\lambda$ to be $0$ and $1$,
therefore the boundaries are located at $\{r(0),\theta(0)\}$ and $\{r(1),\theta(1)\}$. 

Now we consider a sequence of static solutions, varying 
the location of the boundaries. By this variation, the energy changes as 
\bea
 \delta E & = & \left[
  {\partial {\mathfrak L} \over\partial\left({dr\over d\lambda}\right)}
   \delta r +
  {\partial {\mathfrak L}\over\partial\left({d\theta\over d\lambda}\right)}
   \delta \theta
   \right]_0^1 
   \label{deltae}\\
 &\equiv& \left[
      {\tilde{\sigma}(r\sin\theta)^{p-1}\over
        \sqrt{\left({dr\over d\lambda}\right)^2
     +r^2 f \left({d\theta\over d\lambda}\right)^2}
       }\left(
       {dr\over d\lambda}\delta r
       +r^2 f{d\theta\over d\lambda}\delta \theta
        \right)\right]_0^1\,. \nonumber
\eea
The previous equation relates the the difference in energy between two
neighboring static configurations to the `energy flux' at the inner and
outer boundaries. 
The bulk contribution to the difference in energy disappears with the
aid of the equations of motion. 

Here we are interested in the sequence of solutions that describes the
escape of black hole from the brane as shown in
Fig.~\ref{fig1}. For these configurations the inner boundary is then
fixed on the horizon or on the axis at $\theta=0$, while the outer
boundary is lifted, thus describing the recoil in the center of mass
frame of the black hole. 

\begin{center}
\begin{figure}[h]
\hspace*{-7mm}
\scalebox{0.4} {\includegraphics{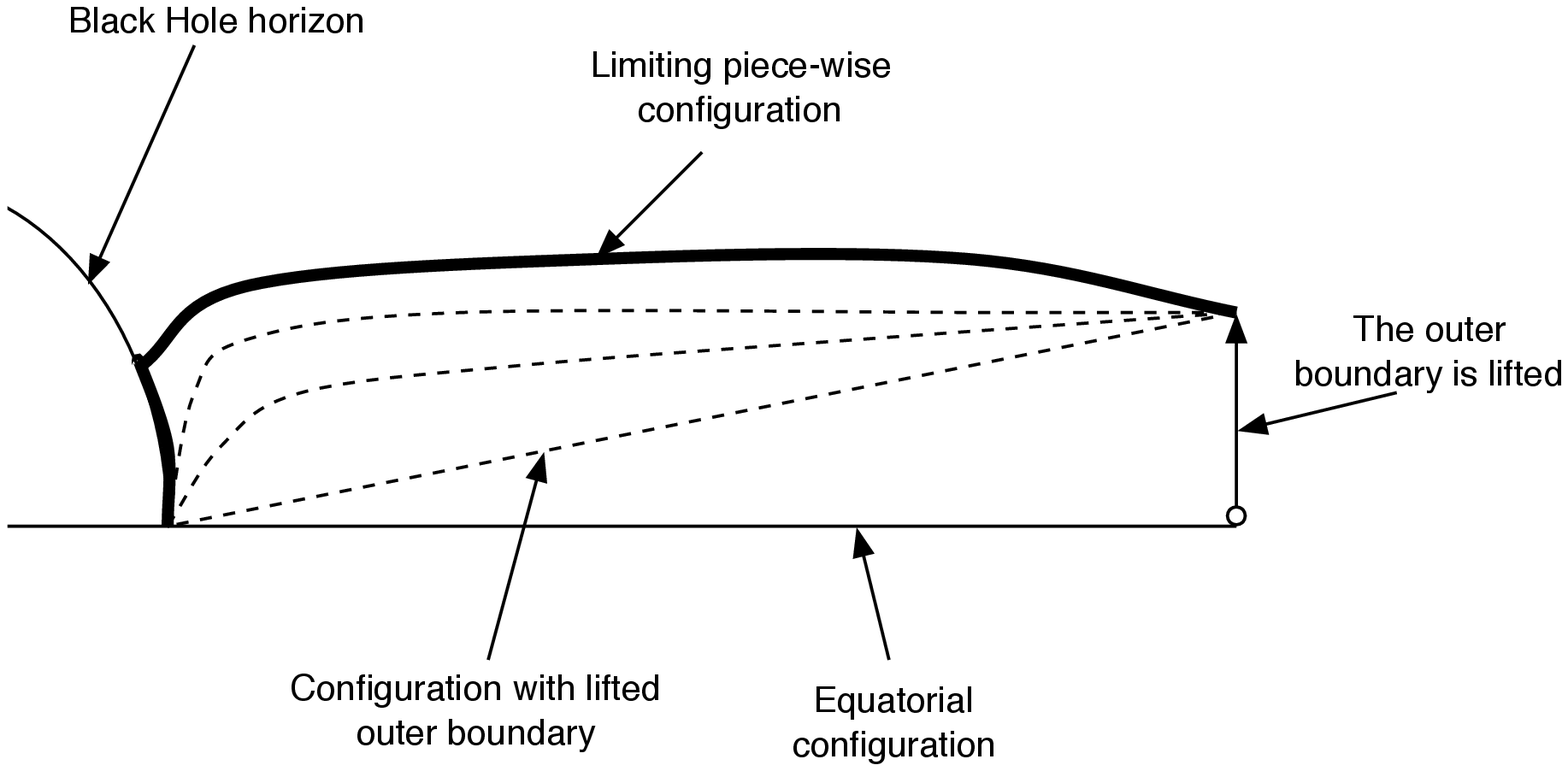}}
\caption{When the outer boundary is slightly lifted the initial
 configuration relaxed to the one with the least action which is the
 piece-wise limiting configuration drawn above.}
\label{fig1-1}
\end{figure}
\end{center}

In estimating whether there is a potential barrier in the path of escape for the black hole, first we have to fix the sequence of configurations.
Thus, given an initial static configuration, we lift the outer boundary of an small amount and look for the next static configuration to which the brane will relax. 
Let us consider the equatorial configuration as the initial one and lift the
outer boundary, as shown in Fig.~\ref{fig1-1}. We know that such an
intermediate configuration with lifted outer boundary does not exist as a stationary one. 
If we introduce a dissipation term to the dynamics, 
the brane will continue to relax to a configuration which realizes 
a smaller value of energy. By minimizing numerically the action with 
fixed inner and outer boundaries, one
can see that the brane evolves towards the limiting
piece-wise configuration presented in Fig.~\ref{fig1-1}, 
which is the one with smallest action amongst
all the configurations with the same boundary conditions. 
This configuration is composed of two pieces: one is the static solution 
which satisfies the regularity condition on the inner boundary and the 
other is the piece lying on the black hole horizon. 
To reach such
a configuration it will take infinite time because of the small lapse 
near the horizon in Schwarzschild coordinates. In the quasi-static
evolution, however, the lifting proceeds through a sequence of
infinitesimal changes of location of the outer boundary. 
Therefore the piece-wise configuration will be
achieved in a good approximation, except for the 
very vicinity of the horizon. 
The contribution 
to the energy from the piece near the horizon is negligibly small
when the configuration is close to the limiting piece-wise one. 
(Notice that $f=0$ on the horizon, and $ dr/d\lambda \approx 0$ 
near the horizon.) 
Now, the portion of the brane lying on the horizon does not contribute to the energy, therefore, as far as energy is concerned, 
the path for the escape will be described by the configurations plotted in Fig.~\ref{fig1}, in which the location of the inner boundary is also varied.
Let us now examine the change rate of the energy of the static
configurations coming from the variation of the locations of 
the inner and outer boundaries. 
First we consider the inner boundary. 
If we take into account the piece lying on the horizon, 
we can say that the inner boundary is always fixed at the 
initial location. 
Therefore its contribution should trivially vanish. 
As mentioned above, we can 
consider the sequence of static solutions instead of 
the piece-wise limiting configurations.  
In this case the inner boundary is varied according to the 
change of the outer boundary. 
However, when the inner boundary is on the horizon, 
we have $\delta r_0=0$ and $f_0=0$, where the subscript $0$ denotes the
value calculated at $\lambda=0$. 
When the inner boundary is on the axis, 
$\delta\theta =0$ and $dr/d\theta=0$. 
Therefore according to the above
expression (\ref{deltae}) the contribution to the difference in energy
from the inner boundary vanishes: $\delta E_0=0$.

More complicated is the evaluation of the contribution coming from the
outer boundary, $\delta E_1$. 
The total energy is divergent if we consider an infinitely extended 
object.  
As a regularization scheme, 
we cut the configurations at a finite radius. 
It is convenient to introduce cylindrical coordinates 
\begin{equation}
 R=r\sin\theta,\qquad
 Z=r\cos\theta~.
 \label{cil}
\end{equation}
For later use, we record the expression of $d\theta/dr$ in terms of 
$dZ/dR$: 
\begin{equation}
 {d\theta\over dr}
 ={1-\tan\theta{dZ\over dR}\over r\left(
    {dZ\over dR}+\tan\theta\right)}~.
\label{dthetadr}
\end{equation}
We consider the situation in which the outer boundary is varied 
with $R$ fixed. Since 
$R$ represents the circumferential radius, 
fixing $R$ provides a natural way to choose the 
boundary in a coordinate invariant manner. 
One could imagine an alternative cutoff radius by fixing the value of $f^\nu R$ with an 
arbitrary power $\nu$. Here we discuss only the simplest case with $\nu=0$, but 
it is easy to check that any choice of the cutoff radius of this kind leads to 
the same conclusions. In this sense, the results obtained 
below are robust. In Appendix we give a little more rigorous 
justification of this chosen prescription for a regularization scheme.

Since the value of $R$ is fixed at the outer boundary, 
$\delta r_1$ and $\delta\theta_1$ are related to each other by 
the relation
\begin{equation}
 \delta R_1=\sin\theta_1 \delta r_1+r_1\cos\theta_1 \delta\theta_1=0, 
\end{equation}
where the subscript $1$ denotes the value at $\lambda=1$. The variation of the energy, $\delta E_1$, due to displacement of the outer boundary can be written as 
\begin{equation}
\delta E_1=\left[
 {\tilde{\sigma} (r\sin\theta)^{p-1} \cos\theta
      \over \sqrt{1+r^2 f \left({d\theta\over dr}\right)^2}}
 \left(1-{rf\tan\theta}{d\theta\over dr}\right)
   \right]_1 \delta Z_1~.\nonumber 
\end{equation}

In the following, we assume that $Z$ becomes almost constant at a large
$R$. This is expected because the spacetime is asymptotically
Minkowski. 
This assumption can be directly verified by numerical integration of
the equation of motion, and it is visually clear already from
Fig.~\ref{fig1}. Under this assumption, we have $dZ/dR\ll 1$ 
and $\tan\theta=R/Z\gg 1$. 
Thus we find that 
the first term in parentheses in the denominator of the R.H.S. of Eq.~(\ref{dthetadr}) 
can be safely neglected. 
At a large distance we can make use of the Newtonian
approximation. Namely, we can treat 
$\delta f$ as a small perturbation denoting $f$ by $1+\delta f$.
Using these approximations, we have 
\begin{equation}
 \delta E_1\approx 
  \tilde{\sigma}\left[R^{-q} Z+R^{p-1} {dZ\over dR}\right]_1\delta Z_1, 
  \label{eq:9}
\end{equation}
where $q=d-(p+1)$ represents the co-dimension of the brane. 
For any type of brane the first term vanishes in the limit 
$R_1\to \infty$. Hence we have 
\begin{equation}
 \delta E_1\mathop{\longrightarrow}_{R_1\to \infty} 
   \tilde{\sigma} R_1^{p-1} \left({dZ\over dR}\right)_1
          \delta Z_1. 
\label{eq:deltaE1}
\end{equation}
If we assume that $\delta E_1/\delta Z_1$ becomes a constant 
independent of $R_1$ at a large distance, we obtain 
$dZ/ dR\to \tilde{\sigma}{}^{-1} R^{1-p}(\delta E_1/\delta Z_1)$.  
Then we have the asymptotic behaviour of $Z$ as 
\begin{equation}
 Z=Z_{\infty}-{R^{2-p}\over (p-2)\tilde{\sigma}}{\delta E_1\over \delta
  Z_1},  
\label{eq:solZ}
\end{equation}
for $p\geq 3$. 
$Z_\infty$ is the asymptotic value of $Z$. 
In the case with $p=2$, $Z$ continues 
to change logarithmically even at a large radius. 
Here we concentrate on the cases with $p\geq 3$. 

It is easy to see from numerical calculations 
and also it is possible to check analytically 
that the asymptotic form of the solution is indeed the one 
mentioned above. In fact, if we write down 
the equation of motion keeping the terms linear 
in $Z$ up to $O(\delta f)$, we obtain 
\begin{equation}
 {d^2Z\over dR^2}+{p-1\over R}{dZ\over dR}=(q-1){Z\over R^{d-1}},
\label{eq:Zeq}
\end{equation}
where we neglected the terms of $O(\delta f)$ multiplied by 
the derivative of $Z$ because $Z$ is almost constant. 
The right hand side gives the leading order correction 
due to $\delta f$. 
we can easily solve the above equation
completely neglecting the right hand side as
$Z=Z_\infty+\alpha R^{2-p}$, where $Z_{\infty}$ and 
$\alpha$ are integration constants. This solution is 
completely consistent with the expression~(\ref{eq:solZ}). 
We can iteratively evaluate the effect due to the 
right hand side of Eq.(\ref{eq:Zeq}). Substituting $Z=Z_\infty$ 
on the right hand side, the correction to $Z$ is obtained as
${q-1\over q(d-3)}Z_\infty R^{2-p-q}$. 
Therefore this correction becomes sub-dominant compared with 
the term $\alpha R^{2-p}$ for a sufficiently large $R$. 
Hence, the use of expression~(\ref{eq:solZ}) is justified.  

\section{Energy of Minkowski-type brane at a large distance}
\label{sec:Minkowski}
We now consider the energy of static Minkowski-type brane solutions. 
As we mentioned in the preceding section, 
we can use Newtonian approximation 
at large $r$. In other words we expand the solution $Z(R)$ in terms of
$\delta f$ as 
$$
Z(R)=Z_{(0)}+Z_{(1)}+\cdots~.
$$
The lowest order solution is the one in Minkowski spacetime and it is
therefore given by $Z_{(0)}=$ constant. $Z_{(1)}$ is the correction of 
$O(\delta f)$. 
The expression for the energy $E$ can be also expanded with respect to $\delta f$ 
formally as 
$$
E=E_{(0)}+E_{(1)}+\cdots~,
$$ 
where 
\begin{eqnarray}
 E_{(0)} & = & \tilde{\sigma}\int_0^1 d\lambda\, (r\sin\theta)^{p-1}
    \sqrt{\left({dr\over d\lambda}\right)^2
      +r^2\left({d\theta\over d\lambda}\right)^2}~,\nonumber \\
 E_{(1)} & = & {\tilde{\sigma}\over 2} \int_0^1 d\lambda\, (r\sin\theta)^{p-1}
    {r^2 {d\theta\over d\lambda}
      \over \sqrt{\left({dr\over d\lambda}\right)^2
      +r^2\left({d\theta\over d\lambda}\right)^2}}\delta f. \nonumber
\end{eqnarray}
Then, we have
\begin{equation}
 E[Z]=E_{(0)}[Z_{(0)}]+\left(\int d\lambda\, {\delta E_{(0)}\over \delta Z} Z_{(1)}
    \right)
   +E_{(1)}[Z_{(0)}]~, \nonumber
\end{equation}
where we have neglected terms of $O\left((\delta f)^2\right)$.
Since the second term vanishes because $Z_{(0)}=$ constant satisfies 
$\delta E_{(0)}/\delta Z=0$, 
we only need to evaluate $E_{(1)}[Z_{(0)}]$ as 
the lowest order correction of $O(\delta f)$. 
Substituting $\delta f=-1/r^{d-3}$ and 
$r\cos\theta=Z_{(0)}$, we have 
\bea
E_{(1)}[Z_{(0)}]
 &=&-\tilde{\sigma} Z_{(0)}^{2-q}\int d\theta (\cos\theta)^{q-1}(\sin\theta)^{p-1}\nonumber \\
 &&\mathop{\longrightarrow}_{R_1\to\infty} -\tilde{\sigma} Z_{(0)}^{2-q}
  {\Gamma\left[p/2\right]\Gamma\left[q/2\right]
    \over 4\Gamma\left[(p+q)/2\right]}. 
\label{E1Z0}
\eea
For any number of co-dimensions $q\geq 1$, 
the $\theta$-integral converges in the limit $R_1\to\infty$. 
The behaviour of $E_{(1)}[Z_{(0)}]$ is peculiar for 
the cases with $q=1$ and $q=2$. 
In the case with $q=1$, $E_{(1)}[Z_{(0)}]$ continues 
to increase for an increasing value of $Z_{(0)}$. 
In the case with $q=2$, $E_{(1)}[Z_{(0)}]$ 
converges to a 
constant value $\tilde{\sigma}/2p$. In other cases, 
$E_{(1)}[Z_{(0)}]$ vanishes in the limit $Z_{(0)}\to \infty$.  

We can now compare the energy of this asymptotic configuration 
with the one on the equatorial plane. 
The difference $E_{(0)}[Z_{(0)}]-E_{\rm equatorial}$ 
is given by the area of the brane cut by the black hole horizon, 
$\tilde{\sigma}/p$. Hence, neglecting the terms of $O((\delta f)^2)$, 
we have 
\bea
 \Delta E \approx
   \left\{
 \begin{array}{ll}
   \displaystyle 
    -\tilde{\sigma} Z_{\infty}{\sqrt{\pi}\Gamma[p/2]\over 4\Gamma[(p+1)/2]}
 & \mbox{for}~ q=1~,\cr
   \displaystyle {\tilde{\sigma}\over 2p}
 & \mbox{for}~ q=2~,\cr
  \displaystyle{\tilde{\sigma}\over p} & \mbox{for}~ q\geq 3~,
 \end{array}\right.
\label{eq:limitE}
\eea
for a large value of $Z_\infty$, 
where we have defined 
\beq
\Delta E\equiv E[Z_\infty]-E_{\rm equatorial}~.
\label{dede}
\eeq
A way of interpreting the previous results is to write down the equation for the gravitational field:
$$
\Box h_{\mu\nu} = T_{\mu\nu} - {1\over d-2} g_{\mu\nu} T~,
$$
where $T_{\mu\nu}$ is the stress tensor coming from the brane. By taking the $00$ component  we obtain an equation for the gravitational potential $h_{00}$:
$$
\Box h_{00} = {q-2\over 2(d-2)}T_{00}~.
$$
The coefficient in the previous expression describes what happens: for $q=1$ the sign is negative and the potential is repulsive, for $q>2$ the sign is positive and the potential is attractive. The case of $q=2$ is marginal.

These results are completely in harmony with those 
obtained in the opposite approximation in which the 
black hole is treated as a test particle on the background spacetime determined by the gravity of the brane.  
In this approximation, co-dimension one branes cannot be
embedded in an asymptotically flat spacetime. The simplest 
solution with the highest symmetry is the Vilenkin-Ipser-Sikivie model. We prepare two copies of Minkowski spacetime and 
consider the inside of the hyperboloid whose invariant distance 
from the origin is constant. Placing a brane on this hyperboloid means 
gluing together two copies of Minkowski spacetime there. 
Since the bulk is simply given by Minkowski spacetime
in this model, the trajectory of a test particle is 
just a straight line. On the other hand, the brane moves 
along a hyperboloid, and hence the brane is always accelerated 
outward. To the contrary, if we look at a test particle 
moving in the bulk from the brane, it is accelerated 
in the direction leaving from the brane. Namely, the force 
acting on the particle looks repulsive. 

When we consider a co-dimension two brane, we can recall 
a cosmic string in four dimensional spacetime, which we know 
is described by Minkowski spacetime with a deficit angle. 
For co-dimension two branes in general dimensions, 
the same is true. Therefore a particle moving far from the 
brane does not feel the gravity force due to the brane at all. 
This is consistent with the fact that the derivative of the 
energy $dE_{(1)}[Z_{(0)}]/dZ_{(0)}$ computed from Eq.~(\ref{E1Z0}) 
vanishes for $q=2$. 
For higher co-dimensions, gravity force caused by a brane 
is attractive and is proportional to $\ell^{1-q}$, where $\ell$ is 
the distance between the brane and the particle. 
This dependence on $\ell$ also agrees with the estimate from 
the derivative of the energy~(\ref{E1Z0}). 

\section{potential barrier for the escape of black hole}
\label{sec:barrier}
\begin{figure*}[t]
\scalebox{0.7} {\includegraphics{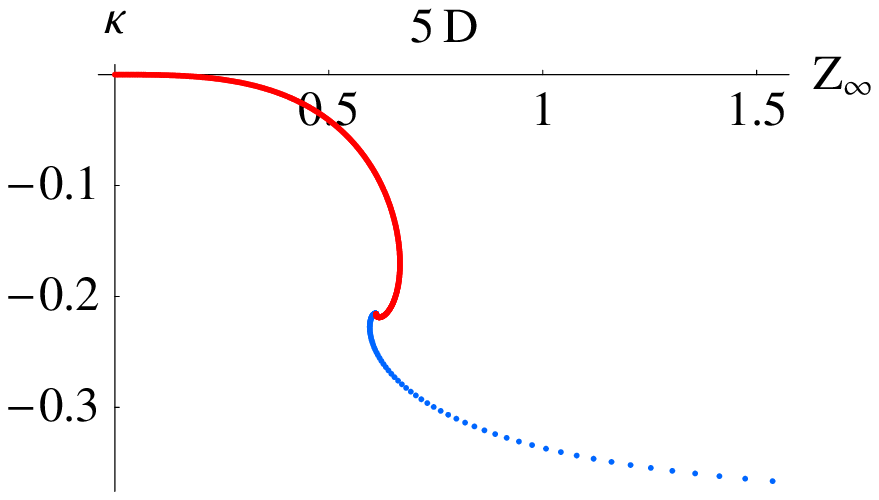}}
\scalebox{0.7} {\includegraphics{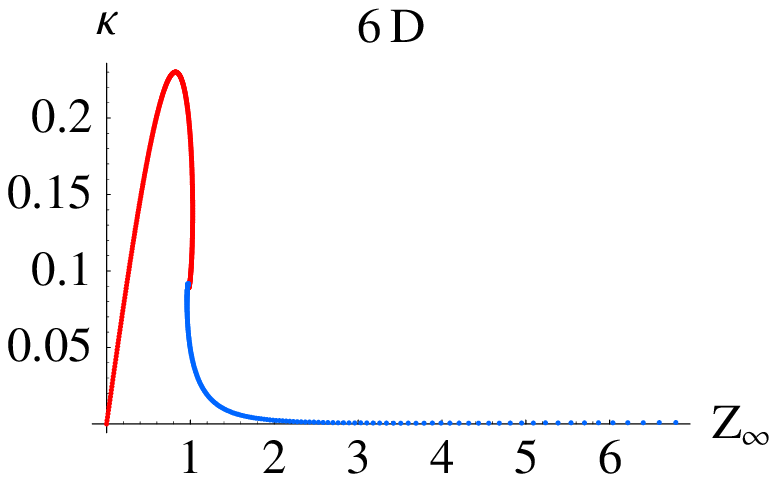}}
\scalebox{0.7} {\includegraphics{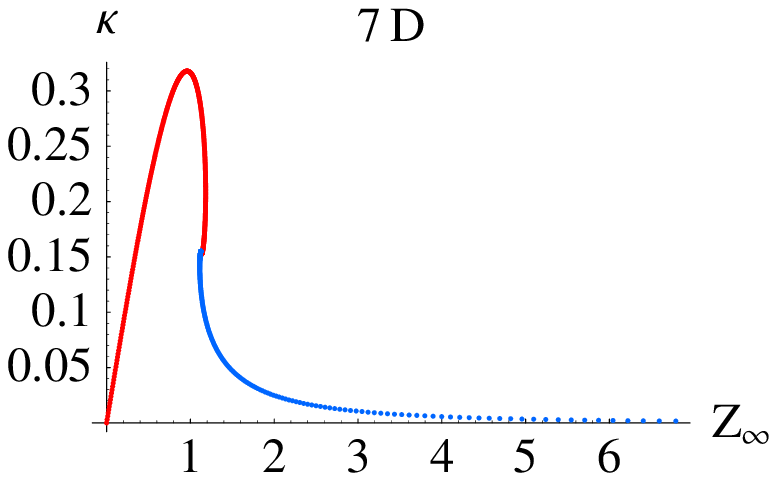}}
\scalebox{0.7} {\includegraphics{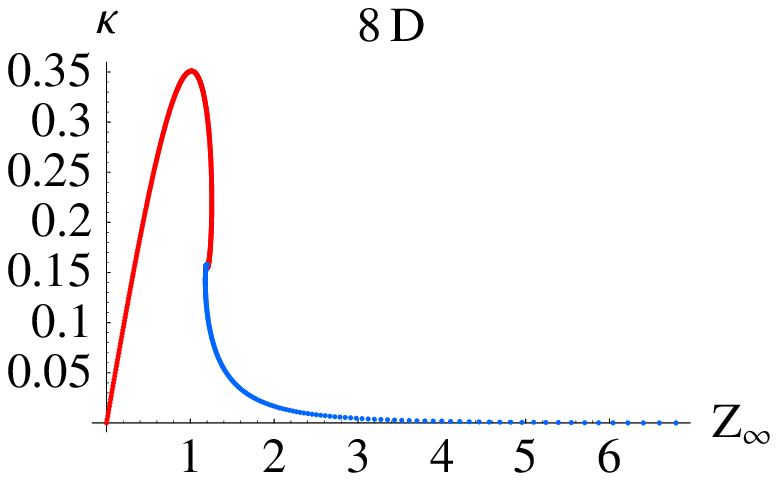}}
\caption{The figure shows the plots of $\{Z_{\infty},\kappa\}$ for a
 $p=3$ brane in a $d = 5, 6, 7, 8$ dimensional bulk respectively. The
 red line corresponds to the brane intersecting the black hole, whereas
 the blue line corresponds to the brane outside the black hole.}
\label{fig2}
\end{figure*}

\begin{figure*}[th]
\scalebox{0.7} {\includegraphics{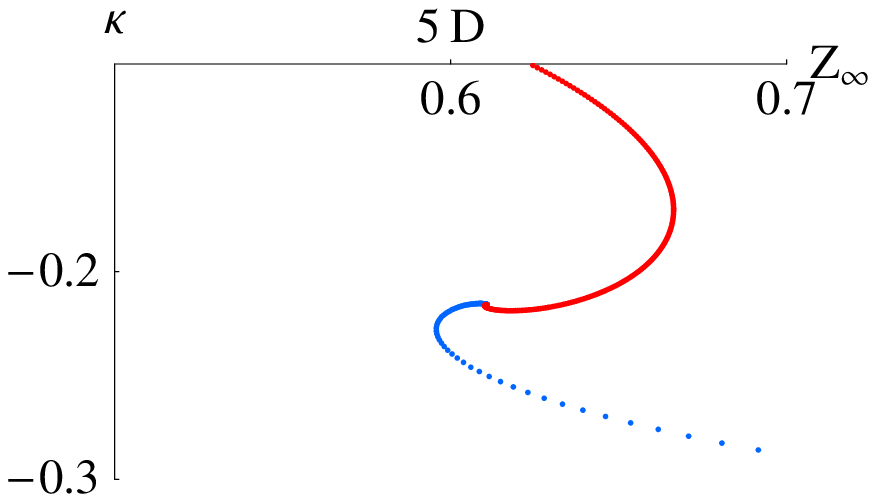}}
\scalebox{0.7} {\includegraphics{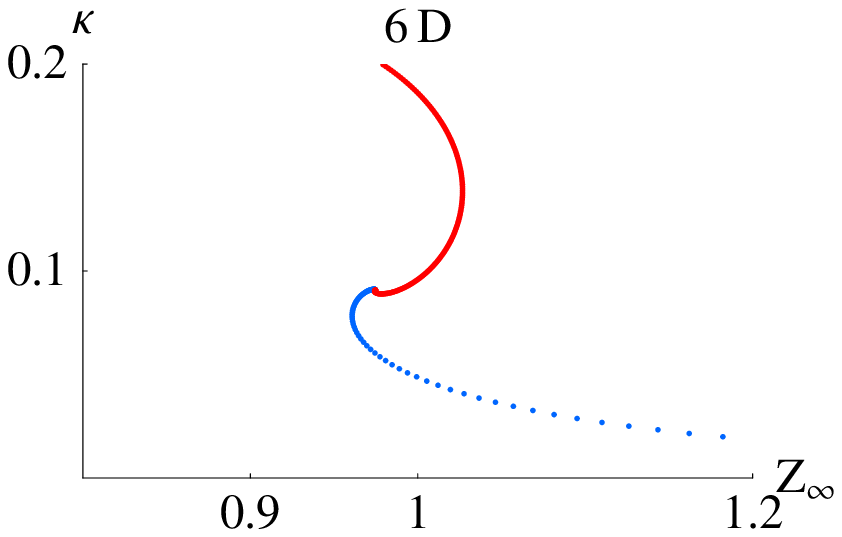}}
\vspace*{5mm}
\scalebox{0.7} {\includegraphics{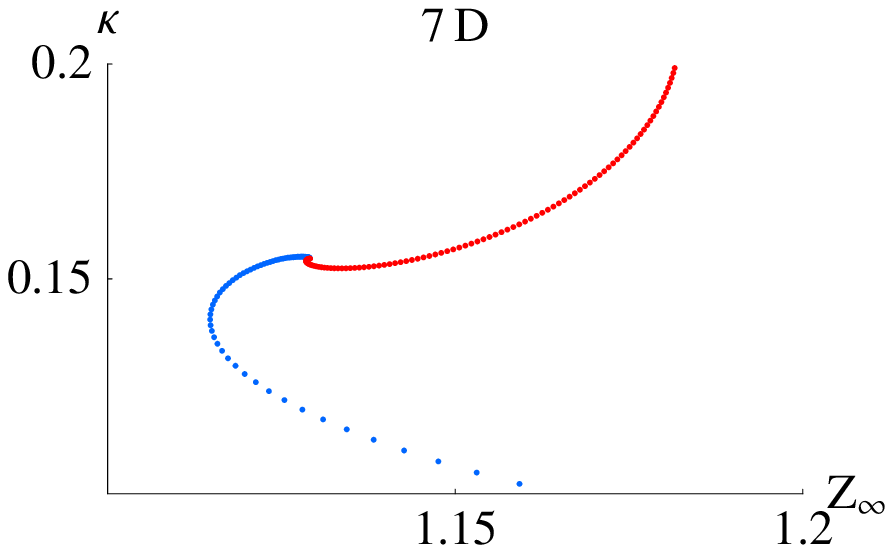}}
\scalebox{0.7} {\includegraphics{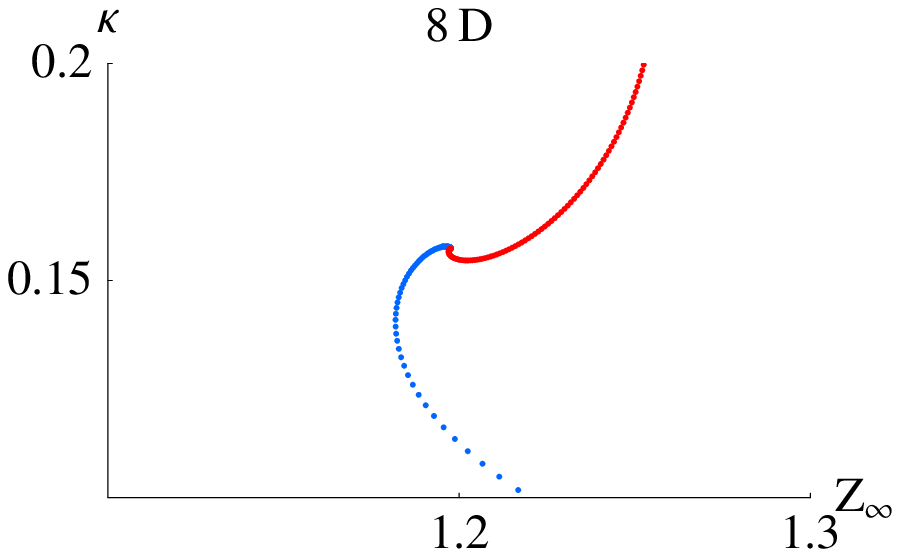}}
\caption{Transition between black hole topology to Minkowski topology.}
\label{fig3}
\end{figure*}

\begin{figure*}[t]
\scalebox{0.8} {\includegraphics{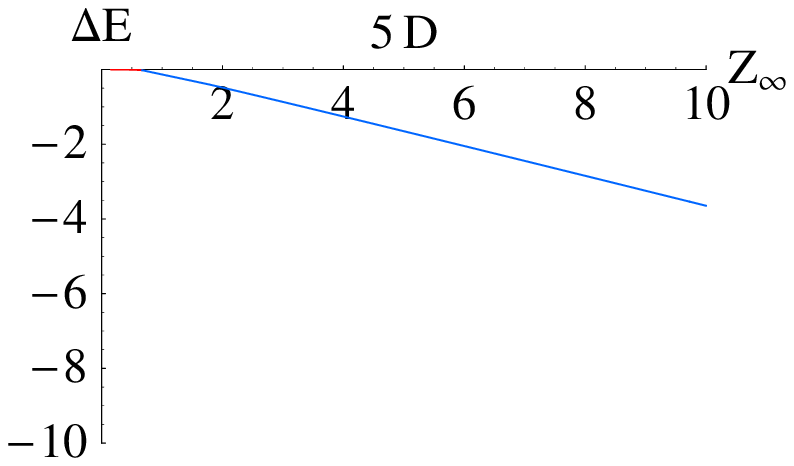}}
\scalebox{0.8} {\includegraphics{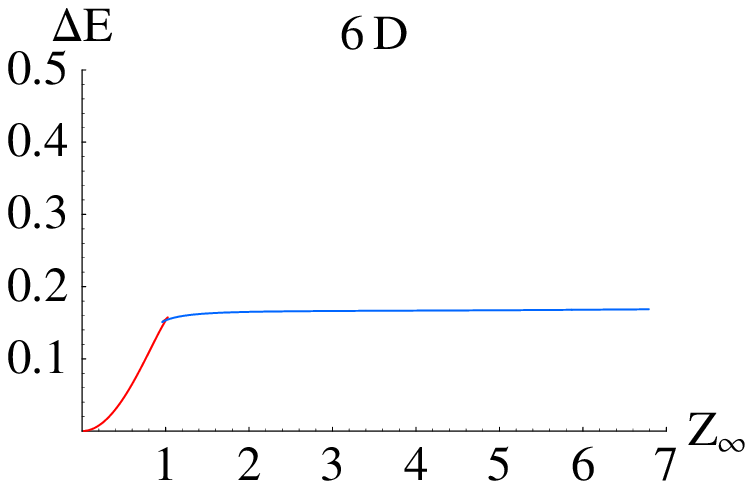}}
\scalebox{0.8} {\includegraphics{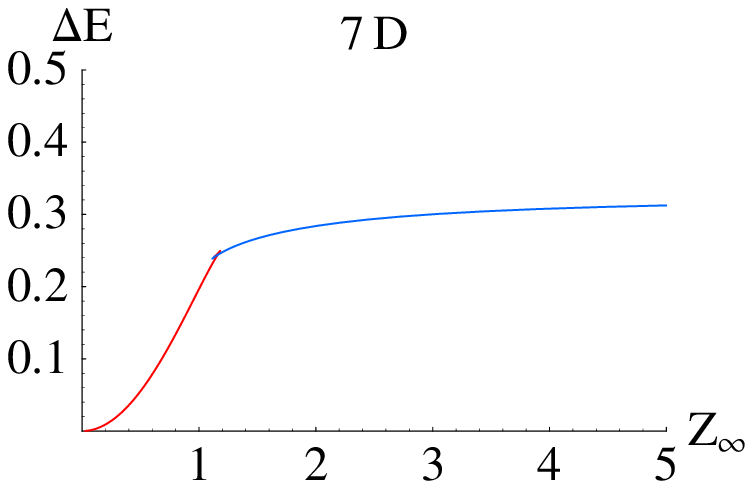}}
\scalebox{0.8} {\includegraphics{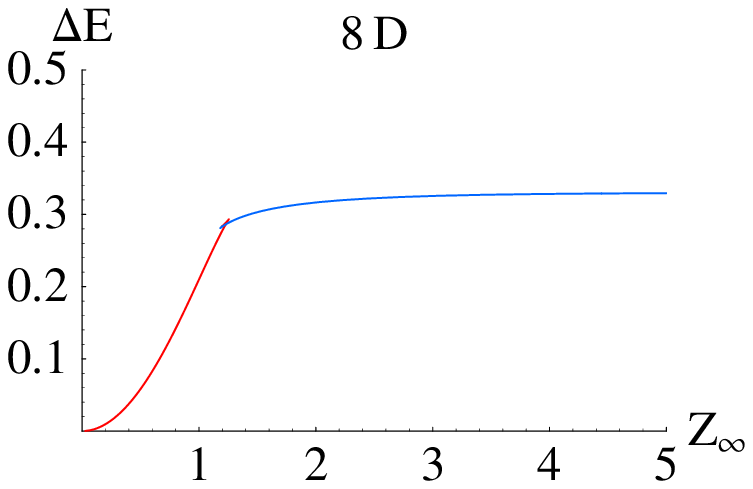}}
\caption{The figure shows the dependence of $\Delta E$ on $Z_\infty$ for $d= 5, 6, 7, 8$ and $p=3$.}
\label{fig4}
\end{figure*}

\begin{figure}[th]
\scalebox{0.8} {\includegraphics{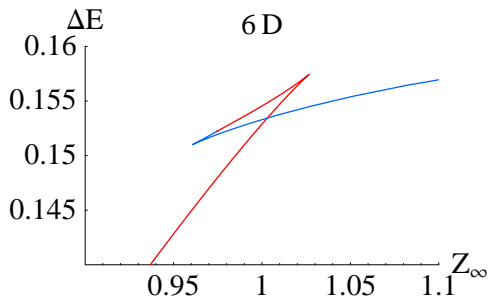}}
\caption{Close up of the profile in the region near pinching. The red branch represents the brane configurations attached to the black hole; the blue branch represents the brane separated from the black hole. The plot refers to the case $d=6$, $p=3$.}
\label{fig3b}
\end{figure}

\begin{figure}[th]
\scalebox{0.4} {\includegraphics{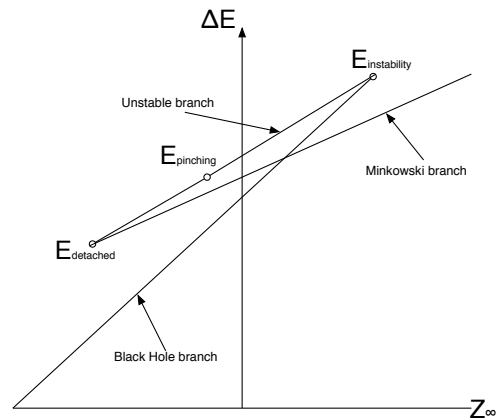}}
\caption{Schematic representation of Fig.~\ref{fig3b}.}
\label{fig3c}
\end{figure}

In the present section we will complement the analytical results of the preceding section by numerical estimates.
To start with, we can compute the profile of the potential along the path of escape of Fig.~\ref{fig1} by calculating the difference of the regularized energy for the sequence of 
static solutions numerically. Here, due to technical simplicity, instead of evaluating the 
energies of the solutions directly, we evaluate at finite radius the 
difference in energy for neighboring configurations 
by using formula (\ref{eq:9}). 
Fig.~\ref{fig2} shows the plots of $\{Z_{\infty},\kappa\}$
with 
\begin{equation}
\kappa \equiv
 {1\over \tilde{\sigma}} {\delta E~\over \delta Z_{\infty}}~,
\end{equation}
for various co-dimensions.
The close-up views of Fig.~\ref{fig3} illustrate the transition point where the brane pinches and goes from a configuration intersecting the black hole to the detached one with Minkowski topology.
Finally, in Fig.~\ref{fig4} we plot the value of the energy $\Delta E$ {\it vs} $Z_\infty$.
Here $\Delta E$, defined by Eq.~(\ref{dede}), is evaluated by the integral $\Delta E=\int
dZ_{\infty} (\delta E/\delta Z_{\infty})$. The asymptotic values of
$\Delta E$ in the limit $Z_{\infty}\to \infty$ are consistent with the
analytic estimates given in Eq.~(\ref{eq:limitE}). 
Fig.~\ref{fig3b} illustrates the detailed structure of the profile for
the energy in the region near the pinching. 
Besides this tiny structure the curves are monotonically decreasing 
for co-dimension one branes and increasing for the other cases. 
The most important point is that there is no energy barrier 
which is higher than the initial value for co-dimension one branes. 
For the other cases with higher co-dimensions, the barrier height  
does not exceed the asymptotic value at $Z_\infty\to\infty$. 
These features can be more easily shown by considering the 
energy of a momentarily static configuration with $Z=$constant.
The energy for this configuration is given by 
\begin{equation}
 E_{Z=\rm const.}=\tilde{\sigma}\int dR R^{p-1}\sqrt{
     1-Z^2(R^2+Z^2)^{-(d-1)/2}}.\nonumber
\end{equation}
The plots of this simple expression are similar to but slightly 
bigger than the results shown in Fig.~\ref{fig4}. 
This simple expression gives an upper bound on the 
potential obtained by correctly solving the equation of motion for 
static configurations. Therefore this 
simplified potential is sufficient to show that there is 
no significantly high bump in the profile of the potential. 
In Fig.~\ref{fig3b} we can distinguish three branches describing the
evolution of the brane: the (stable) black hole branch where the brane intersects
the black hole horizon, the (stable) Minkowski branch where the 
brane lies outside the black hole horizon and finally 
an unstable branch joining the first two. 
This last unstable branch contains 
the configuration attached to the pole at $\theta =0$ on the event horizon, which 
we call the pinching configuration. 
This configuration represents a kind of `geometric
threshold' between the configurations that intersect the black hole and
those that do not. 
If we closely look at the vicinity of the pinching configuration, 
there is a finer structure~\cite{art1}, 
which is beyond the resolution of the plot given in Fig.~\ref{fig3b}. 
Schematically, these three branches 
are drawn in Fig.~\ref{fig3c}.
Accordingly, we can distinguish three energy scales: $E_{pinching}$,
$E_{detached}$ and $E_{instability}$. $E_{pinching}$ labels the energy
of the pinching configuration. 
The values of $E_{pinching}$ are tabulated in Table 1
for all dimensions and co-dimensions up to $d=11$.
\begin{center}
\begin{table}[h]
\begin{tabular}{|c|c|c|c|c|c|c|}
        \hline
$p\diagdown d$ &  $6$    &   $7$  & $8$  &   $9$  & $10$ &   $11$     \\
        \hline
 3  & 0.1523 & 0.2405 & 0.2830 & 0.3044 & 0.3158 & 0.3223    \\
        \hline
4  & - & 0.1171 & 0.1800 & 0.2099 & 0.2254 & 0.2340   \\
        \hline
5  & - & - & 0.0962 & 0.1477 & 0.1706 & 0.1822 \\
        \hline
6  & - & - & - & 0.0820 &  0.1251  & 0.1436    \\
        \hline
7  & - & - & - & - &  0.0687  & 0.1081    \\
        \hline
8  & - & - & - & - &  -  & 0.0608    \\
        \hline
\end{tabular}
\caption{Values of $E_{pinching}$ for $d=6, ..., 11$ and $p=3, ..., 8$.}
\end{table}
\end{center}
$E_{detached}$ represents the minimum energy of a configuration detached
from the black hole. It is interesting to note that such `detached'
configuration cannot be reached through a sequence of static solutions
starting from the initial equatorial configuration, but nevertheless it
marks the threshold for the separation between the brane of the black
hole, which can in principle be reached via some different (dynamical)
deformation of the brane.
Finally, $E_{instability}$ represents the value of the energy
corresponding to the upper cusp (see Fig.~\ref{fig3c}), which marks an onset of instability. If the brane starts from the initial equatorial
configuration and goes through the sequence of static solution it will
reach such `instability' point. The question is what happens when this
instability point is passed over. If we introduce dissipation we can expect that the brane will evolve towards a configuration with lower energy lying on the Minkowski
branch and therefore a configuration separated from the black hole will
be realized. As one can notice from Fig.~\ref{fig4}, the difference
between the values of $E_{pinching}$, $E_{detached}$ and
$E_{instability}$ is very small. Some values of $E_{detached}$ and
$E_{instability}$ are tabulated in Table 2 for $p=3$ and $d=6, ..., 11$.
\begin{center}
\begin{table}[h]
\begin{tabular}{|c|c|c|c|c|c|c|}
        \hline
$d$ &  $6$    &   $7$  & $8$  &   $9$  & $10$ &   $11$     \\
        \hline
 $E_{instability}$  & 0.1574 & 0.2491 & 0.2926 & 0.3145 & 0.3264 & 0.3333 \\
        \hline
 $E_{detached}$  & 0.1510& 0.2384 & 0.2806 & 0.3019 & 0.3134 & 0.3200    \\
        \hline
\end{tabular}
\caption{$E_{instability}$ and $E_{detached}$ for $p=3$ and $d=6, ..., 11$}
\end{table}
\end{center}

As anticipated, we find that the co-dimension one brane is peculiar in
the sense that it does not have a barrier for the escape of black hole. 
This means that there is no critical escape velocity in the present
approximation where the brane tension is not very large. 
By contrast, all other types of branes have a barrier. Roughly speaking,
the height of the energy barrier is 
$O(\sigma (G_d m)^{p/(d-3)})$ where we have explicitly written the $d-$dimensional Newton constant
$G_d$ and the black hole mass $m$. 
Then, by equating the kinetic energy of the black hole $m v^2/2$ with this energy barrier, 
the critical escape velocity $v_c$ will be estimated as $v_c\approx \sqrt{\sigma G_d^{p/(d-3)} m^{(2-q)/(d-3)}}$. Very interestingly, the critical escape velocity $v_c$ is independent of the mass of the black hole for co-dimension two branes ($q=2$).  In the cases with higher co-dimensions $v_c$ is smaller for a black hole with a larger mass, as is expected from the larger inertia of the black hole.  

\section{discussion}
\label{sec:discussion}
In this paper we investigated a system composed of a brane and a black
hole when a perturbation gives the black hole a relative velocity with
respect to the brane. The principal problem we are interested in is the
phenomena of escape of the black hole. 
The main goal of our study is to clarify whether the effect of the brane
tension may prevent the black hole from escaping for small recoil
velocities and to understand whether a critical escape velocity exists. 

Here, we analyzed this problem by studying the interaction between a
Dirac-Nambu-Goto brane and a black hole assuming adiabatic
(quasi-static) evolution. 
By `quasi-static' we mean that the system evolves very slowly going
through configurations which are stationary. Taking the brane lying on
the equatorial plane of the black hole as initial configuration, we
considered the brane in a fixed Schwarzschild spacetime background,
neglecting the gravitational perturbation caused by the brane,
approximation which is adequate in the case of small tension.
Our strategy is to compute the energy for the sequence of the static
solutions describing the escape as shown in Fig.~\ref{fig1} directly and
see how the energy changes along the easiest path for the black hole to
escape. From this knowledge we can estimate in which cases there is a
potential barrier that prevents the separation and the initial energy
necessary to overcome such a barrier.

We gave an analytic estimate for the height of the potential by
comparing the energies of the initial and the final configurations,
where `final configuration' refers to the one with the black hole
disconnected and being far away from the brane. We also computed the
energy of each static configuration numerically, thus providing support
to the analytical estimates.
From the calculated shape of the potential, we find that there is no
barrier in the potential for
co-dimension one branes, meaning that the configuration with a black
hole attached to a 
co-dimension one brane is unstable. In contrast, when the co-dimension
of the brane is equal to or greater than two, there is a potential
barrier along the escape path of the black hole, and the critical escape
velocity is evaluated to be $O(\sqrt{\sigma G_d^{(d-q-1))/(d-3)}
m^{(2-q)/(d-3)}})$, where $d$ and $q$ are the spacetime dimension and
brane co-dimension, respectively. 

In Ref.~\cite{art3}, we studied the phenomena of escape treating the
brane as a domain wall in a scalar effective field theory. 
Since here we have focused on the case Dirac-Nambu-Goto branes, it is
interesting to ask whether the results of this paper are also valid when
the brane has a thickness. We expect that even in the case of thick
branes described by field theoretical topological defects, the analytic
comparison of the energies between the initial and the final state will
not change dramatically. 
For the final configuration the description by a Dirac-Nambu-Goto brane
is always a good approximation. For the initial configuration, the
contribution to the energy from the region near the horizon may be
corrected, especially when the thickness of the defect is not very small
compared with the horizon radius. However, the height of the barrier in
the intermediate stage is potentially quite different, 
and therefore a more rigorous study
of the thick case is necessary to make a definite statement.

As we mentioned in the introduction, one of the main motivations that
led us to consider the problem of how a black hole 
interacts with a domain wall is related to the possibility 
of observing mini black holes in
collider experiments or in cosmic rays. In this context Ref.~\cite{stoj}
points out that the escape of the black hole might provide a way of
distinguishing between various models. Specifically, 
one possible difference might come out from the fact that, contrary to
models with large extra dimensions \cite{r1,r1-1}, warped models have additional
$Z_2-$symmetry \cite{r2}, which makes the brane behave as if the tension were
infinite, resulting in the impossibility for mini black holes to
leave the brane. This is, however, not the case in models when the
$Z_2-$symmetry is relaxed, as Refs.~\cite{art1,art2,art3} have
shown. The results presented here take a further step in this direction
and suggest that also within models of the same class, without
$Z_2-$symmetry, the co-dimensionality has a remarkable effect. If the
co-dimension is $1$, then even for small recoil velocity the black hole
is expected to slide off the brane, contrary to the higher co-dimension
case where there is an energy barrier that would prevent this from 
occurring.

At this point it seems interesting to compare the recoil velocity due to Hawking radiation 
to the critical velocity. If we assume that the recoil is due to
inhomogeneous Hawking emission, a rough estimate gives $v_{rec}\simeq
1/\sqrt{N}$, where $N \simeq  m/T_H$ is the number of emitted
particles and $T_H\approx r_H^{-1}$ is the Hawking temperature. 
The horizon radius $r_H$ is given by $(m/M_d^{d-2})^{1\over d-3}$
with $M_d$ being the higher dimensional Planck scale. 
Thus, we have
\beq
v_{rec}/v_{cr} \simeq  \left({M_d\over m}\right)^{1+p\over 2d-6} \left({M_d^{1+p}\over \sigma}\right)^{1/2}~,
\label{rr}
\eeq
From this equation, we can compute the critical value for the mass at
which the recoil velocity equals the critical velocity:
$$
m_{cr} \simeq {M_d^{d-2}\over \sigma^{d-3\over p+1}}~.
$$
Thus we can say that, if the initial black hole mass is smaller than the
critical mass, the recoil velocity is expected to
exceed the critical one before long, and therefore 
the black hole will leave the brane during the early  
stage of its evaporation. On the other
hand, if the mass of the black hole is larger than $m_{cr}$, 
we expect that the black hole will escape from 
the brane in a subsequent stage after its mass becomes as small as 
$m_{cr}$ after losing a significant portion of its initial mass.

Finally we want to mention the validity range of the probe-brane
approximation.  
The probe-brane approximation will require that the mass of 
the portion of the brane near the black hole horizon is 
much smaller than the black hole mass:
$$
\sigma r_H^p \ll m \,. 
$$
This gives a constraint 
$$
 m \gg M_d \left({\sigma\over M_d^{p+1}}\right)^{{d-3\over q-2}}~.
$$
This constraint is not so stringent. 
As far as the brane tension is small 
in the higher dimensional Planck units, 
our approximation will be kept to be valid throughout the
above-mentioned process 
of black hole evaporation until its escape from the brane at 
$m\approx m_{cr}$. 

\acknowledgements This work is supported in part by Grant-in-Aid for Scientific
Research, Nos. 1604724 and 16740141 and the Japan-U.K. Research Cooperative Program
both from Japan Society for Promotion of Science. This work is also supported by the 21st Century COE ``Center for Diversity and Universality in Physics'' at Kyoto university, 
from the Ministry of Education, Culture, Sports, Science and Technology of Japan and by Monbukagakusho Grant-in-Aid for Scientific Research(S) No. 14102004 and (B) No.~17340075. A.F. is supported by the JSPS under contract No. P047724. 

\appendix
\section{justification of the regularization procedure}
\label{sec:justification}
We present one argument for justification of using the 
cutoff at a fixed circumferential radius. 
One way to make the energy well defined is to fix the 
outer boundary configuration completely; not only the 
position of the brane but also the gravitational field. 
More precisely, for example, fixing the boundary 
induced metric will be appropriate. 
This is impossible 
in the present setup since the spacetime is exactly 
given by Schwarzschild spacetime.   
If we fix the location of the brane at the boundary, 
a different static solution corresponds to 
a different position of the black hole. 
Then, the induced metric on the boundary changes. 
To fix the boundary metric, we need to include metric 
perturbations. 

For clarity, we can consider the Newtonian picture. 
We set the outer boundary at $r=r_{o}$ and 
the brane is on the equatorial plane on the boundary.  
Then, the perturbation of the Newtonian potential $\phi$
necessary 
to compensate the change of the location of the central black hole
will be evaluated as 
\begin{equation}
  \delta \phi\approx {d-3\over 2}
   {Z\over r_{o}^{d-1}}\delta Z. 
\end{equation}
Here $G_d m$ is set to unity as before. 
The leading order perturbation of the gravitational potential 
will be given by a dipole field  
$\delta \phi ={d-3\over 2}
   {Z\delta Z\over r_o^{d-1}}$, 
which scales 
as $\sim 1/r_o^{d-1}$ for large $r_o$. 
The total amount of correction due to this perturbation 
of the gravitational potential to the 
gravitational binding energy of the brane will 
scale like $\delta\phi$ multiplied by 
the area of the brane $\sim r_o^p$. Hence it scales as 
$\sim 1/r_o^{q}$. We can therefore neglect this contribution 
in the limit $r_o\to \infty$. 
The correction to the energy due to 
gravitational perturbations is quadratic in $\delta\phi$. 
Hence, its contribution scales as 
$\delta\phi^2\times r^{d-1}\propto 1/r_o^{d-1}$, 
and we can neglect it safely. 
When we take into account the gravity, we also need to consider 
gravitational potential induced by the brane. However, 
the background gravitational potential is chosen to be a 
vacuum solution. Then the leading order correction is quadratic. 
Therefore it scales like $\sigma^2$. 
In the end, we claim that the effect of gravity 
can be completely neglected when the outer boundary radius $r_o$ 
is moved to infinity as far as we consider energy of $O(\sigma)$.

\end{document}